# Charged liquid surface instabilities and detection of Solar neutrino by XENONnT

Sergey Pereverzev
Lawrence Livermore National Laboratory

E. Aprile *et al.* [1] report the detection of solar $^8$B neutrinos in the 5.9-ton active volume XENONnT dual-phase detector—an important milestone for microphysics and background studies in dark matter detectors. Reference [1] claims rejection of the background-only hypothesis with a statistical significance of 2.73σ. But, "anomalous responses" are noted in [1–3]:
(i) enhanced calibration signals near wire crossings for perpendicular wires added to the gate and anode grids for rigidity;
(ii) anomalous S2 shapes at these locations, revealed by machine learning analyses [1,4];
(iii) a sharp rise in single-electron emission with extraction field, limiting extraction to ~50%.

The data analysis introduces constants $g_1$ and $g_2$—light and charge amplification parameters (in detected photons) [1–3]. To account for local electric field variations affecting electron extraction and S2 electroluminescence, varying correction coefficients are applied [1–3]. In the absence of explanations for (i)–(iii), data from "anomalous regions" (Fig. 1a) were excluded from analysis. We argue that the continuous PTFE wall surrounding the active surface impedes removal of unextracted charges, leading to surface charge accumulation, which is known to cause multiple surface instabilities. Suppression of electron extraction by surface charge is known [5] and can cause systematic errors in $g_2$.

Delayed bursts of electron emission ("e-bursts") appear at S2 locations of prior high-energy events (PHE) in XENON1T and LUX (which has only two PTFE wall openings) [6]. Such bursts were seen in XENON10, XENON100, and other detectors, but not in RED1, ZEPLIN-III, or RED100—where designs facilitate surface charge removal [7–9]. In [12], we argued that e-bursts arise from charged surface instabilities, including formation of a lattice of surface hillocks [10] with characteristic length scale $\lambda = 2\pi a$, where $a = (\sigma/\rho g)^{1/2}$ is the capillary length. In LHe, $a \approx 2.5$ mm, and a few mm in LXe. In stronger fields, hillocks can evolve into charged droplets or jets [11]. The hillock lattice forms in uniform fields; XENONnT wire spacing is close to $a$, and the field distribution may override the instability pattern to follow the wire layout (Fig. 1a). Field inhomogeneity of $a$ scale was used to define hillock locations in electrospraying experiments [11] (Fig. 1b).

Thus, wire crossings represent local maxima in charge density and charged surface elevation. Anomalies (i)–(iii) likely arise from collective interactions among electrons, negative ions, surface wave resonances, and surface profiles—phenomena long known to produce instabilities [12] and complex behavior [13]. Since surface charge density increase is not confined to wire crossings, excluding these regions does not eliminate systematic errors due to suppressed electron extraction.

The standard extraction model assumes electron overheating in strong fields [5,15,16], enabling electrons to overcome the ~0.7 eV potential barrier at the liquid surface [5]. Electron drift proceeds via elastic scattering on Xe atoms [16]. However, surface charges can introduce inelastic processes, transferring energy to surface-bound electrons or capillary waves, trapping outgoing electrons.

Early LXe extraction studies found that brief field shutdown restored extraction [5]. Later measurements, including [15,17], applied vetoes after energetic events to allow surface charge relaxation. Different studies show extraction saturation (kinks) at different fields [15], underscoring the need for calibration across all energy regions.

For smaller detectors (XENON10, XENON100, ZEPLIN-III), a background rising toward low energy was reported in combined S1-S2 and S1-only analyses; similar excesses appear in other detectors [18]. S1-S2 XENON1T analysis shows a comparable low-energy excess [19]. XENONnT does not exhibit a rising low-energy background [20], though delayed electron and photon emission from PHEs should be present. In XENON1T, $^{37}$Ar L-shell (2.8 keV) decays were used in S2-only calibrations to cross-check S1-S2 analyses [21]. XENONnT's lowest-energy calibration used only $^{37}$Ar K-shell S1-S2 coincidence, with no report (to our knowledge) of a K-L branching ratio check.

In RED-100 (~100 kg LXe), operated near the Kalinin reactor, delayed 2–5 electron events exceeded expected reactor antineutrino signals by ~two orders of magnitude, independent of reactor status [9]. While RED-100 experiences higher muon fluxes than deep underground detectors, improved charge removal in XENONnT might increase low-energy event detection in both S1-S2 and S2-only analyses [2].

We conclude that significant, unquantified systematic uncertainties may affect $g_2$ estimates for small signals in XENONnT. Targeted experimental studies are urgently needed to evaluate surface charging effects on small-signal detection.

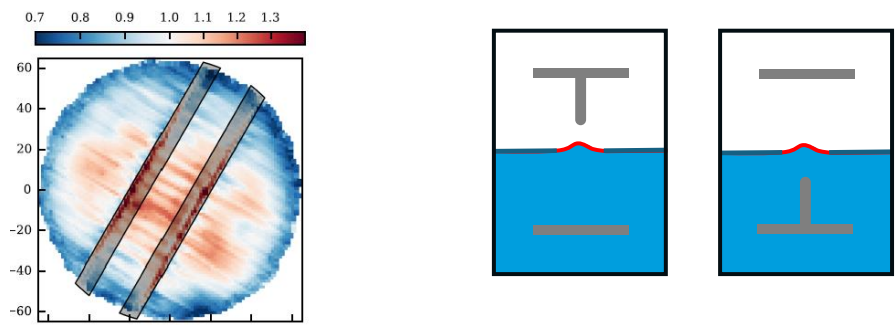

Fig.1 a -S2 corrections map reflects pattern of grid wires, figure from [3], grey area data were excluded from analysis. 1 b- re-drawn based on [15]; position of hillock fixed by rounded electrode with radius ~ *a*, same surface deformation and charge accumulation patterns for tip below and above liquid.

## References

1. E. Aprile, et al., First Indication of Solar $^8$B Neutrinos via Coherent Elastic Neutrino-Nucleus Scattering with XENONnT, Phys. Rev. Lett. 133, 191002 (2024)   DOI: 10.1103/PhysRevLett.133.191002

2. E. Aprile, et al., Search for New Physics in Electronic Recoil Data from XENONnT Phys. Rev. Lett. 129, 161805 (2022)   DOI: 10.1103/PhysRevLett.129.161805
3. E. Aprile et al., XENONnT analysis: Signal reconstruction, calibration, and event, Phys, Rev, D 111, 062006 (2025)    DOI: 10.1103/PhysRevD.111.062006
4. Sophia Farrell, an enhanced approach to signal analysis in the XENONnT dark matter experiment, a dissertation submitted in partial satisfaction of the requirements for the degree of Doctor of Philosophy In physics, Rice University, May2024
5. Gushchin et al., Electron emission from condensed noble gases, Sov. Phys. JETP 49(5), pp. 857-858 (1979)
6. D.S. Akerib, et al., Investigation of background electron emission in the LUX detector, Phys, Rev, D 102, 092004 (2020)   DOI: 10.1103/PhysRevD.102.092004
7. D.Y. Akimov et al., Observation of delayed electron emission in a two-phase liquid xenon detector. JINST 11, C03007 (2016).
8. D.Y. Akimov et al., The ZEPLIN-III dark matter detector: instrument design, manufacture and commissioning, Astropart. *Phys.* **27** (2007) 46 [astro-ph/0605500].
9. D.Y. Akimov et al., First constraints on the coherent elastic scattering of reactor antineutrinos off xenon nuclei, Phys. Rev. D 111, 072012 (2025)
10. S. Pereverzev, What surfaces in operation of noble liquids dark matter detectors, *JINST* 18 C07011 (2023)  **DOI** 10.1088/1748-0221/18/07/C07011
11. P. Leiderer, *Electrons at the surface of quantum systems*, J. Low Temp. Phys. **87** (1992) 247-278   https://doi.org/10.1007/BF00114906.
12. P. Moroshkin, P. Leiderer, T.B. Möller and K. Kono, Taylor cone and electrospraying at a free surface of superfluid helium charged from below, Phys. Rev. E **95** (2017) 053110 https://doi.org/10.1103/PhysRevE.95.053110
13. Landau, L. D. and E. M. Lifshitz: Fluid Mechanics, Pergamon Press, 1959
14. Boyle, F.P., Dahm, A.J. Extraction of charged droplets from charged surfaces of liquid dielectrics. *J Low Temp Phys* **23**, 477–486 (1976). https://doi.org/10.1007/BF00116935
15. J. Xu, S. Pereverzev, B. Lenardo, J. Kingston, D. Naim, A. Bernstein, K. Kazkaz, and M. Tripathi, Electron extraction efficiency study for dual-phase xenon dark matter experiments. Phys. Rev. D 99, 103024 (2019)   doi.org/10.1103/PhysRevD.99.103024
16. M. H. Cohen and J. Lekner, Theory of Hot Electrons in Gases, Liquids, and Solids, Phys. Rev. 158, 305 (1967), DOI: https://doi.org/10.1103/PhysRev.158.305
17. T. Pershing, et al., Measurement of the ionization yield from nuclear recoils in liquid xenon between 0.3--6 keV with single-ionization-electron sensitivity, Phys. Rev. D **106**, 052013 (2022),  DOI: https://doi.org/10.1103/PhysRevD.106.052013
18. S. Pereverzev, Detecting low-energy interactions and the effects of energy accumulation in materials, Phys. Rev. D **105** (2022) 063002    https://doi.org/10.1103/PhysRevD.105.063002
19. E. Aprile, et al., Excess electronic recoil events in XENON1T, Phys. Rev. D 102, 072004 (2020), DOI: 10.1103/PhysRevD.102
20. Mateo Rini, Potential Dark Matter Signal Gives Way to New Limits, Physics Magazine 15, 159 (2022)   DOI: 10.1103/Physics.15.159
21. E. Aprile, at al., Low-energy calibration of XENON1T with an internal [37]Ar source, Eur. Phys. J. C (2023) 83:542        https://doi.org/10.1140/epjc/s10052-023-11512-z